\documentclass[a4paper,twocolumn, nofootinbib,superscriptaddress,aps,prl,eqsecnum,notitlepage,showkeys]{revtex4-1}
\usepackage{multirow}
\usepackage{graphicx}
\usepackage{bm}
\usepackage{array}
\usepackage{dcolumn}
\usepackage{hyperref}
\usepackage{gensymb}
\usepackage{amsmath}
\usepackage{dcolumn}    
\usepackage{ifpdf}
\usepackage{amssymb,lineno,amsfonts}
\usepackage{graphicx}   
\usepackage{bm}         
\usepackage{bbm}
\usepackage{mathrsfs}
\usepackage{upgreek}
\usepackage{mathtools}
\usepackage{epstopdf}
\usepackage{setspace}
\usepackage{hyperref}
\usepackage{natbib}
\usepackage{multirow}
\usepackage{esvect}
\usepackage{soul}
\usepackage{amsmath}
\usepackage[usenames,dvipsnames]{xcolor}
\definecolor{med-blue}{RGB}{25,25,112}
\hypersetup{colorlinks, linkcolor={blue},citecolor={Blue}, urlcolor={blue}}
\usepackage{times}
\usepackage [english]{babel}
\usepackage [autostyle, english = american]{csquotes}
\MakeOuterQuote{"}
\begin{document}
\title{Anisotropy in the magnetization and magnetoelectric response of single crystalline Mn$_{4}$Ta$_{2}$O$_{9}$  }
\author{Soumendra Nath Panja}
\affiliation{Department of Physics, Indian Institute of Science Education and Research \\ Dr. Homi Bhabha Road, Pune, Maharashtra-411008, India}
\author{Pascal Manuel}
\affiliation{ISIS Pulsed Neutron Source, STFC Rutherford Appleton Laboratory, Didcot,\\ Oxfordshire OX11 0QX, United Kingdom}
\author{Sunil Nair}
\affiliation{Department of Physics, Indian Institute of Science Education and Research \\ Dr. Homi Bhabha Road, Pune, Maharashtra-411008, India}
\affiliation{Centre for Energy Science, Indian Institute of Science Education and Research,
	\\ Dr Homi Bhabha Road, Pune, Maharashtra-411008, India
	}
\date{\today}
\begin{abstract} 
We report temperature dependent magnetization and neutron diffraction measurements on the corundum related magnetoelectric antiferromagnet Mn$_{4}$Ta$_{2}$O$_{9}$. Measurements performed on a single crystalline specimen reveal that the magnetization is anisotropic, and a weak ferromagnetic component emerges well within the antiferromagnetically ordered state. Powder neutron diffraction measurements indicate that the magnetic structure comprises of antiferromagnetically coupled ferromagnetic chains of Mn$^{2+}$ spins aligned along the trigonal $c$ axis, in contrast to that reported in other isostructural members of this family.  Magnetic measurements performed under a periodic electric field indicate that the magnetoelectric response is also anisotropic, with this coupling along the trigonal $c$ axis and that perpendicular to it having different signs. 
\end{abstract}
\pacs{Pacs}
\maketitle
\section{Introduction} 
   Using symmetry considerations, Dzyaloshinskii predicted that Cr${_2}$O${_3}$ would be magneto-electric, implying that in this system the application of an electric field would induce a magnetic moment, and vice versa \cite{Dzyloshinskii}. This was soon verified experimentally, with Astrov demonstrating the measurement of electric field induced magnetization \cite{AstrovCr2O3}, followed by the measurement of magnetic field induced electric polarization by V.J.Folen \textit{et.al} \cite{Folen}. This area rose to the forefront of experimental and theoretical investigations more recently, with the emergence of research in multiferroics - systems with concomitant and spontaneous electric and magnetic orders\cite{KFWang,Fiebig,Eerenstein,KimuraReview}. The linear magnetoelectric effect is expressed as  P$_{i}$=$\alpha_{ij}$ M$_{j}$or M$_{i}$=$\alpha_{ij}$ P$_{j}$, where P is the polarization, M is the magnetization and the proportionality constant $\alpha_{ij}$ quantifies the linear magnetoelectric coupling \cite{Figbig_JPD}. The magnetoelectric susceptibility $\alpha_{ij}$ is a reducible second rank tensor with possible diagonal and off-diagonal elements as obtained from the coupling response parallel and perpendicular to a given crystallographic orientation. The tensor elements are non zero only when space inversion and time reversal symmetry are both broken. Because of such severe symmetry restrictions there are relatively few materials which are linear magnetoelectrics. Prominent examples include Cr$_{2}$O$_{3}$\cite{Cr2O3}, MnTiO$_{3}$\cite{MnTiO3}, NdCrTiO$_{5}$\cite{NdCrTiO5}, LiFePO$_{4}$\cite{LiCoPO4} and  Co$_{4}$Nb$_{2}$O$_{9}$ \cite{ArimaCoNbO}.
   
The corundum derived honeycomb magnets of the form $A_4X_2$O$_9$ ( with $A$ = Mn, Co or Fe and $X$ = Ta or Nb ) crystallizes in the same point group (\textit{ $\overline{3}^{'}m$}) as Cr$_2$O$_3$, and has attracted recent interest owing to its promising magnetoelectric properties \cite{Bertaut, Fischer}. Based on structural considerations alone, S. C. Abrahams had earlier predicted that a specific member of this family, Mn$_4$Ta$_2$O$_9$ (crystallizing in the  $ R\bar{3}c $ space group) could host a potential  ferroelectric state \cite{Abrahams2007}. However, subsequent investigations have revealed that the members of this family are predominantly linear magnetoelectrics crystallizing in the trigonal $P \bar{3}c1$ symmetry \cite{FeNbO, CoTaO, CoNbO_2011, CoNbO_2014, MnTaO, MnNbO}, with the notable exception of Fe$_4$Ta$_2$O$_9$ which is now known to be genuine multiferroic \cite{FeTaO, FeTaO_PRM}.  Unlike Cr$_2$O$_3$, in these systems, the magneto-electric effect can arise from two non equivalent $A$ sites and could have potentially resulted in a giant linear magneto-electric effect if they had contributed additively \cite{solovyev}.  It has been suggested that the absence of spontaneous ferroelectric polarization is due to the balance of two different AFM domains with equal and opposite ME coefficients. In the presence of an external magnetic field, one of the domains is preferred, and results in the appearance of a finite electric polarization. The details of the magnetic structures even in Co$_4$Nb$_2$O$_9$ - the most extensively investigated member of this family - remain contentious, with N.D. Khanh et.al \cite{ArimaCoNbO} and G.Deng et. al\cite{G_Deng} reporting magnetic structures contradictory to each other. The former inferred that the Co$^{2+}$  spins at two crystallographic sites are aligned in the $ab$ plane with a canting angle of  20$\degree$ along the $c$ axis, while the latter suggested that canting is within the $ab$ plane. Though the closely related Mn$_4$Ta$_2$O$_9$ is now known to exhibit electric polarization in the presence of an external magnetic field \cite{Fang}, its magnetic structure remains to be reported. Moreover, low temperature investigations of single crystalline specimens of this system are also lacking, presumably due to the paucity of high quality single crystals. 

Here, we report on the magnetic and electric field dependent magnetic measurements on single crystals of the Mn${_4}$Ta${_2}$O${_9}$ system, with measurements being carried out parallel and perpendicular to the trigonal axis. We have also investigated the magnetic structure using powder neutron diffraction, and our data indicates that the Mn$^{2+}$ spins are aligned along the crystallographic $c$ axis. Our magnetic measurements also indicate the presence of an anisotropic low temperature weak ferromagnetic phase, which presumably arises as a consequence of the antisymmetric Dzyaloshinskii-Moriya interaction.  Using time dependent magnetic measurements performed under a periodic square electric field, we also infer that the magnetoelectric susceptibility is anisotropic, with this coupling being positive along the $c$ axis,  and exhibiting the opposite sign along the $ab$ plane. 

\begin{table}
	\caption{Structural Parameters of Mn${_4}$Ta${_2}$O$_{9}$ as determined from the Rietveld analysis of room temperature X-ray diffraction data.}
	\begin{tabular}{c c c c c c }
		\hline
		\multicolumn{6}{c} {Mn${_4}$Ta${_2}$O$_{9}$ } \\
		\multicolumn{6}{c}{Temperature = 298 K ; Space Group : $P \bar{3}c1$} \\
		\multicolumn{6}{c}{Crystal system: Trigonal }\\
		\multicolumn{6}{c}{ a= 5.31(32) ${\AA}$ ,b=5.31(32) ${\AA}$ ,c= 14.29(33) ${\AA}$}\\
		\multicolumn{6}{c}{${\alpha}$ = ${\beta}$ = 90 \degree ,${\gamma}$ =  120 \degree}\\
		\hline
		Atom & Wyckoff & x/a &y/b & z/c &  occupation \\
		\hline
		Mn$^1$  & 4d & 0.3333 & 0.6667  & 0.01(88)&0.3333 \\
		Mn$^2$ & 4d & 0.3333& 0.6667  &0.30(53)&0.3333\\
		Ta  & 4c & 0 & 0 & 0.35(69) &0.3333\\
		O1 & 6f & 0.27(73) & 0 & 0.2500&0.4999 \\
		O2 & 12g & 0.33(49) & 0.30(79)&0.08(63)&1 \\
		\hline
	\end{tabular}
	\label{Table1}
\end{table}
\begin{figure}
	\centering
	\includegraphics[scale=0.30]{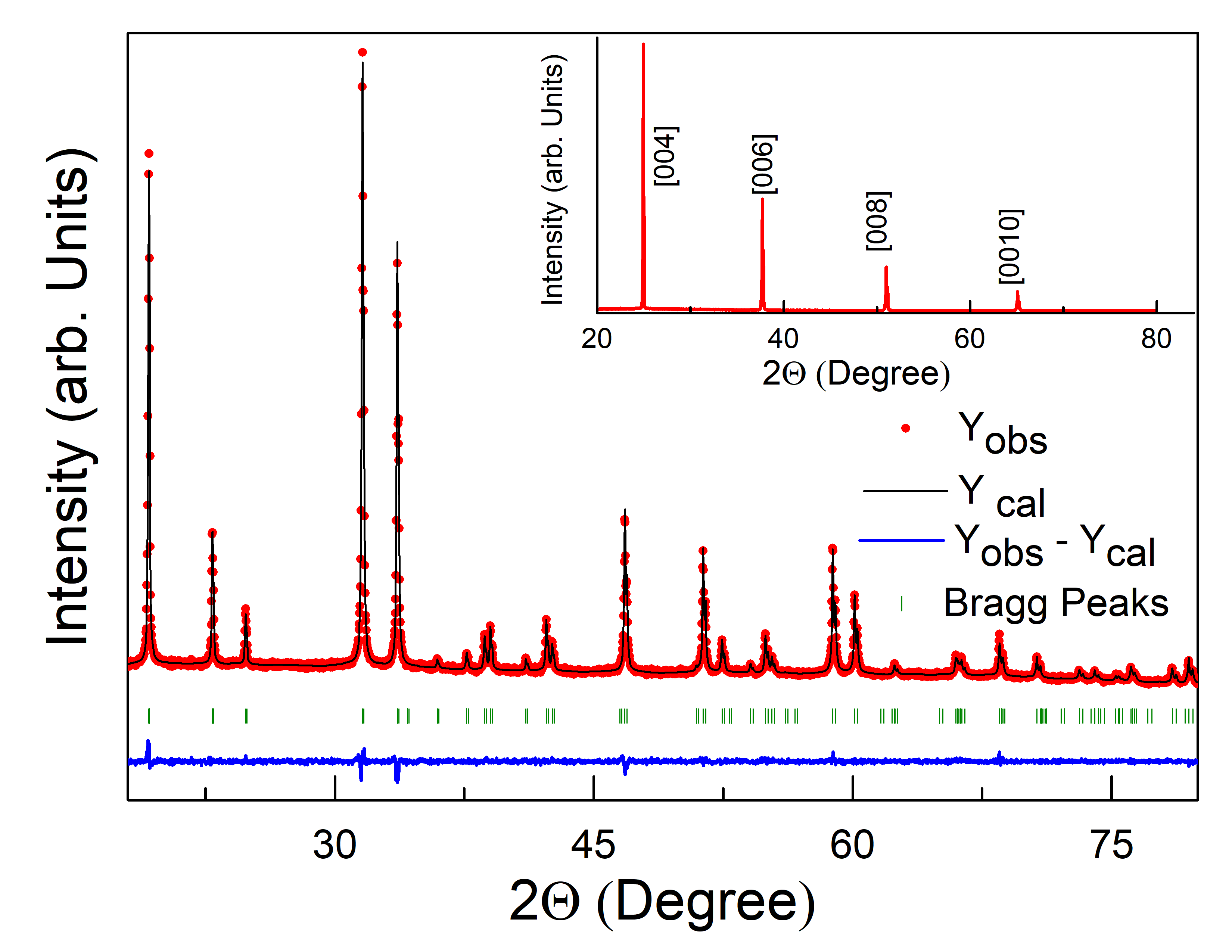}
	\caption{ (Color online ) The main panel depicts Rietveld refinement of room temperature powder X-ray diffraction data of Mn$_4$Ta$_2$O$_9$ .This corresponds to a fit with $R$ parameters of $R{_{wp}} =12.3$, $R{_e} = 9.34$, and $\chi{^2} = 1.72$. The Inset shows the $\theta$-2$\theta$ scan of a representative single crystal specimen.}
	\label{Fig1}
\end{figure}
\section{Experimental} 
Polycrystalline Mn$_{4}$Ta$_{2}$O$_{9}$ was synthesized by the standard solid state reaction route. High purity Mn$_{2}$O$_{3}$  and Ta$_{2}$O$_{5}$ precursors were mixed stoichiometrically and ground using a ball mill for 24 hrs at 120 rpm. This mixture was sintered at 900 $^o$C followed by treatments at 1050  $\degree$C and 1200 $\degree$C, all under argon atmosphere. The final treatment was repeated thrice with intermediate grinding to ensure a homogeneous particle size of the powder. X-ray diffraction was performed on the final polycrystalline powder using Cu K$_{\alpha}$ radiation, and the structure was analyzed by Rietveld refinement using the FULLPROF suite \cite{Fullprof}. Polycrystalline feed and seed rods each of 10 mm length and 7 mm diameter respectively were prepared by  packing the powder inside a rubber tube and applying a hydrostatic pressure of $\approx$ 700 bar. These compact rods were then sintered at 1200$\degree$C under argon atmosphere. The crystal growth was performed using a four mirror optical floating zone furnace ( Crystal System Corporation Japan) with each lamp of 500 watt. At an argon pressure of 1 atm, the growth was unstable and inhomogeneous melting of the feed rod was observed. This was possibly due to the evaporation of Mn from the system, which was further confirmed  by the observation of deposition at the wall of quartz enclosure. The growth was then repeated under an Ar atmosphere of 0.19 MPa, where a stable growth condition was observed. The molten zone was moved vertically down at a rate of 5 mm/h and an axially symmetric counter rotation of both the feed and seed rods at 16 rpm was set in order to maintain temperature and phase homogeneity at the molten zone. Powder diffraction patterns of the frozen melted zone exhibit similar pattern as of the specimen prepared by solid state reaction, indicating that the melting behavior of Mn$_4$Ta$_2$O$_9$ is congruent. Crystal pieces were obtained by cleaving the grown crystal rod. The crystallographic orientations were determined by back reflection Laue X-Ray photography after polishing the crystal pieces. The crystal orientation was  determined by an exact matching of the simulated and observed patterns using the Orient Express software \cite{Oriexpress}. Stoichiometry of the crystal pieces were further determined by energy dispersive x-ray (EDX) from ZEISS Ultra Plus. Temperature and magnetic field dependent magnetic measurements along different crystallographic axes were performed using a Quantum Design MPMS-XL magnetometer. The electric field dependent magnetic measurements were performed using the Manual Insertion Utility Probe of the MPMS-XL magnetometer and a Keithley Source Meter (Model 2612B). Temperature dependent powder neutron diffraction measurements were carried out using the time-of-flight WISH diffractometer at the ISIS neutron facility \cite{TOF_expt}. For solving the magnetic structure, representation analysis was carried out using BasIreps from FULLPROF  and  the SARAh program \cite{ASWills}, and the magnetic structure models were refined using the FULLPROF suite.
\section{Result \& Discussion } 
  \begin{figure}
	\centering
	\includegraphics[scale=0.70]{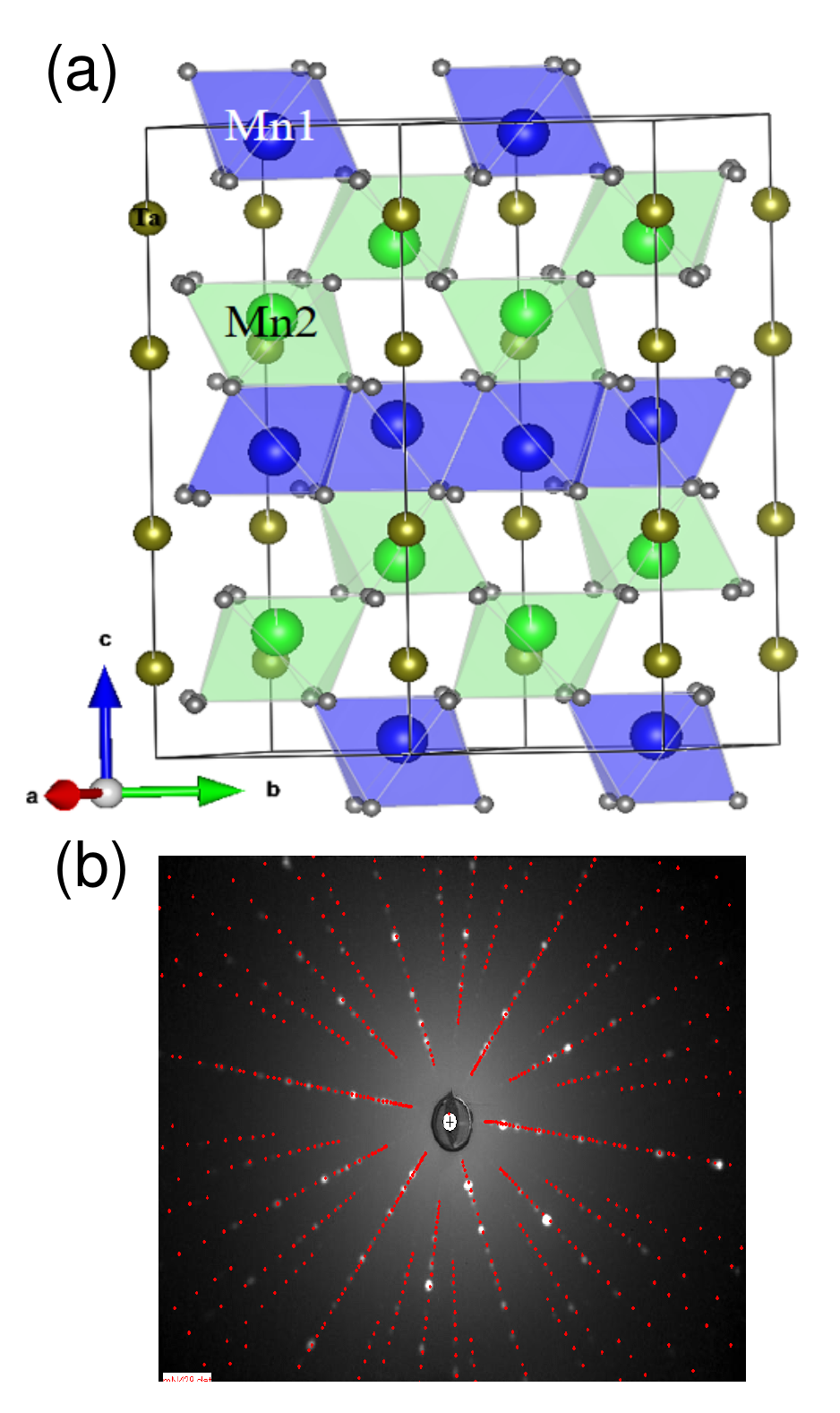}
	\caption{ (Color online ) (a)  Crystal structure as obtained from the  Rietveld refinement of room temperature powder x-ray diffraction data. (b) The fitting of measured Laue pattern with a simulated one using cell parameters deduced from the Rietveld analysis.}
	\label{Fig2}
\end{figure}
 There have been conflicting reports about the structure of Mn$_{4}$Ta$_{2}$O$_{9}$, which is reported to crystallize in both the $ R\bar{3}c $ \cite{GrinsR3c} and the $P \bar{3}c1$ \cite{GrinsP31C} space groups. We find that this system crystallizes in the $P \bar{3}c1$ trigonal space group, and the Rietveld refinement of our  powder x-ray difraction data  is shown in Fig.\ref{Fig1}. The obtained crystallographic parameters are summarized in Table. \ref{Table1}. The inset of Fig.\ref{Fig1} depicts the diffraction pattern obtained from $\theta$-2$\theta$ scan of a representative crystal, and peaks which are solely indexed by [00 2l] planes (with l=2,3,4,5) are observed. A representation of the structure of Mn$_{4}$Ta$_{2}$O$_{9}$ is shown in Fig.\ref{Fig2}(a), with Mn1 and Mn2 denoting the two different crystallographic sites which Mn${^{2+}}$ occupies. The Laue pattern as obtained from back scattered diffraction of one of the cleaved crystals is shown in Fig.\ref{Fig2}(b). Here, the white spots are the observed diffraction spots, and the red dots denote the simulated pattern generated by the Orient Express software \cite{Oriexpress} using the unit cell parameters deduced from our Rietveld analysis. This, along with the absence of any additional spots in the Laue patterns indicate that the Mn$_{4}$Ta$_{2}$O$_{9}$ single crystal is of high quality. A number of crystallites were characterized in this fashion, and all of them were observed to cleave along the $ab$ plane.  Energy dispersive x-ray analysis reconfirmed that the Mn:Ta atomic percentage is in the ratio 2:1. 

\begin{figure}
 	\centering
 	\hspace{-0.5mm}
 	\includegraphics[scale=0.36]{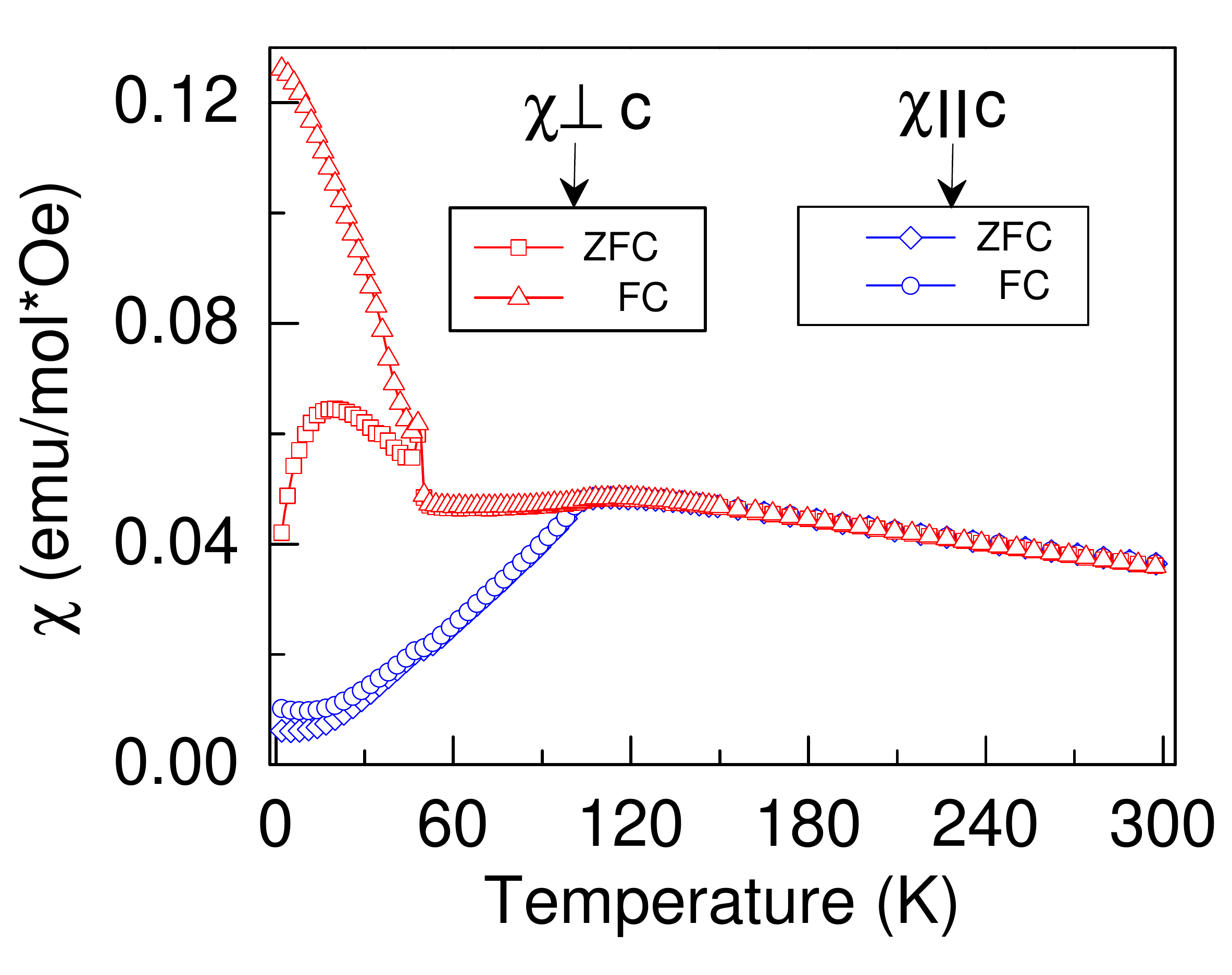}
 	\caption{ (Color online ) Depicts the DC magnetization  of  Mn$_{4}$Ta$_{2}$O$_{9}$ as measured at 100 Oe in the zero field cooled (ZFC) and field cooled (FC) protocols along different orientations of the crystal. }
 	\label{Fig3}
 \end{figure}

 DC magnetic measurements performed in the standard Zero Field Cooled (ZFC) and Field Cooled (FC) protocols for both the crystallographic directions  $H \parallel c$ and $H \perp c$ are depicted in Fig.\ref{Fig3} in the form of the longitudinal susceptibility ($\chi_{\parallel}$) and the transverse susceptibility ($\chi_{\perp}$) respectively.
 A broad anomaly corresponding to the onset of antiferromagnetic order is observed at $T{_N} \approx$ 110K, which is in line with what was reported earlier for polycrystalline specimens of Mn$_{4}$Ta$_{2}$O$_{9}$ \cite{Fang}. Our data reveals that just below the antiferromagnetic transition, the magnetization along the $ab$ plane is relatively temperature independent, whereas that along the crystallographic $c$ axis drops rapidly. This implies that the Mn$^{2+}$ moments are aligned along the crystallographic $c$ axis. Recent measurements on single crystal specimens of Mn${_4}$Nb${_2}$O${_9}$ have exhibited similar signatures \cite{Cao_MnNbO}.  
 \begin{figure}
 	\centering
 	\hspace{-1.0cm}
 	\includegraphics[scale=0.30]{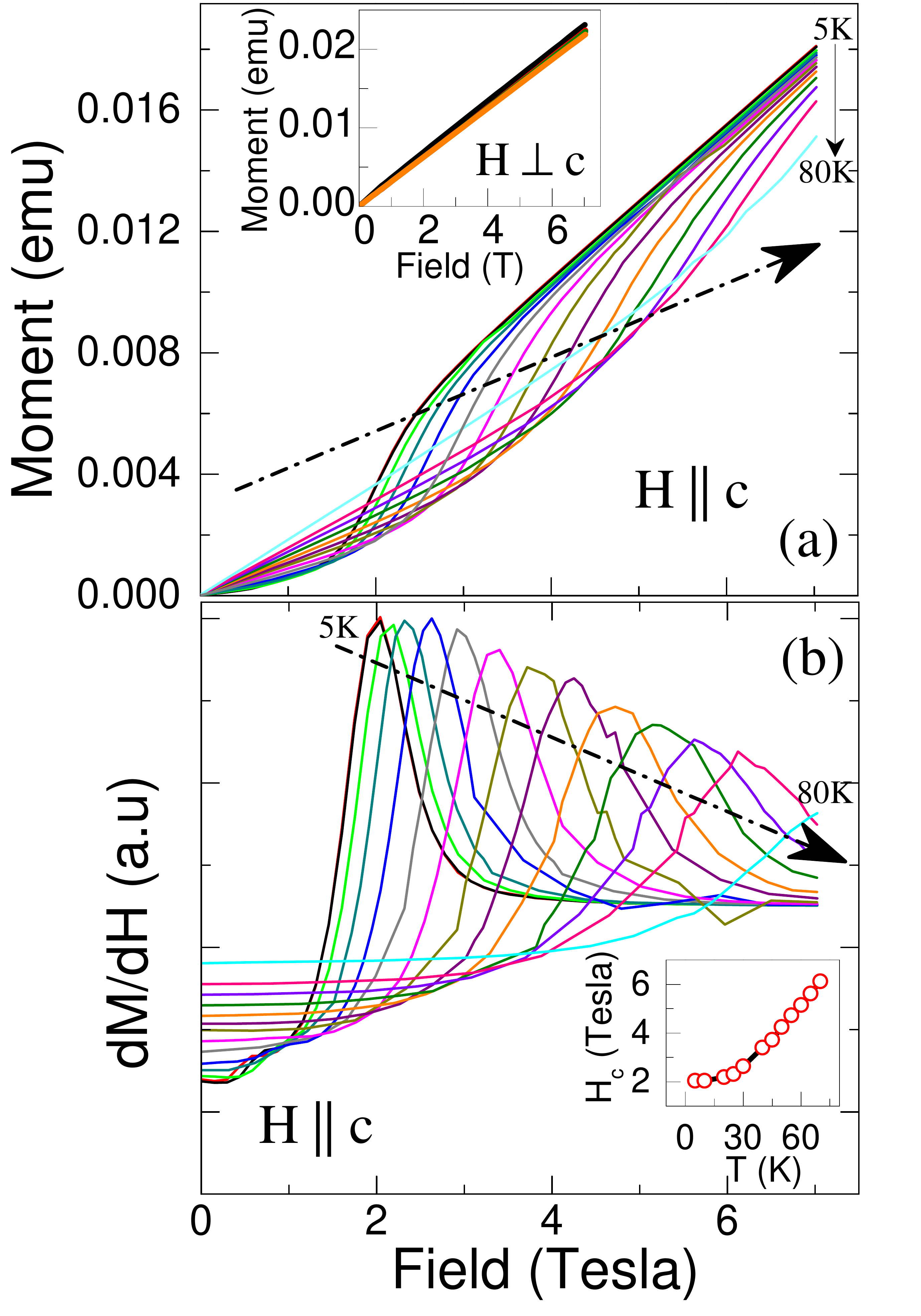}
 	\caption{ (Color on line ) (a) $M-H$ isotherm of  Mn$_{4}$Ta$_{2}$O$_{9}$ as measured along the $H \parallel c$ direction at different temperatures. The inset depicts isotherms measured along the orthogonal $H \perp c$ direction (b) $dM/dH$ of the isotherms measured along $H \parallel c$, with the inset depicting the temperature evolution of the metamagnetic critical field.}
 	\label{Fig4}
 \end{figure}
 Interestingly, we observe the appearance of an additional transition at $\approx$ 48 K in the transverse susceptibility $\chi_{\perp}$ as depicted in Fig. \ref{Fig3}. This feature, which is absent in the longitudinal susceptibility $\chi_{\parallel}$, possibly corresponds to a weak-ferromagnetic phase arising from the well known antisymmetric Dzyaloshinskii-Moriya interaction, making the magnetic phase diagram of Mn${_4}$Ta${_2}$O${_9}$ richer than the other members of this family where no such features have been observed.  Upon further cooling, the perpendicular magnetic susceptibility increases till 19 K and then exhibits a downturn. On the other hand, the longitudinal susceptibility decreases monotonically until 19 K below which it remains relatively temperature independent. This anomalous low temperature behavior of $\chi_{\perp}$ and $\chi_{\parallel}$ below 19K is reminiscent of that observed in single crystals of the brownmillerite  Ca$_2$Fe$_2$O$_5$ \cite{Ca2Fe2O5}, where a regime of anisotropic weak ferromagnetism (below 150K) was observed deep within the antiferromagnetic regime ($T_N$ = 750K). It was also demonstrated that the easy axis associated with this ferromagnetic component could be rotated by the use of a moderate magnetic field. 
 
Magnetic isotherms measured along the two different crystallographic orientations as a function of temperature reveals the presence of a metamagnetic transition. $M-H$ isotherms measured with $H \parallel c$ shows a field induced transition as evidenced by a change of slope [Fig.\ref{Fig4}(a)], whereas this feature is absent in the $H \perp c$ isotherms [inset of Fig.\ref{Fig4}(a)]. This is seen more clearly in curves of $dM/dH$ vs $H$, where this transition manifests itself in the form of a peak [Fig.\ref{Fig4}(b)]. The critical field ($H_{c}$) of this field induced transition varies as $T^2$ [inset of Fig.\ref{Fig4}(b)], as was also observed in the closely related Fe${_4}$Ta${_2}$O${_9}$ system \cite{FeTaO}. Such a  magnetic field driven anomaly can be ascribed to spin flop transition for an AFM where after a critical magnetic field, the  magnetization rotates perpendicular to the magnetic easy axis provided the magneto crystalline anisotropy is weak enough \cite{Neel,Nagamiya}. The observed $T^2$ dependence of $H_c$ is in broad agreement with Yamada’s theory \cite{yamada}, which was formulated to explain field driven phenomena in itinerant metamagnets. However, as in the case of Fe${_4}$Ta${_2}$O${_9}$ and other materials, the applicability of this theory in the case of insulating systems is suspect. 

\begin{figure}
  	\centering
  	\includegraphics[scale=0.40]{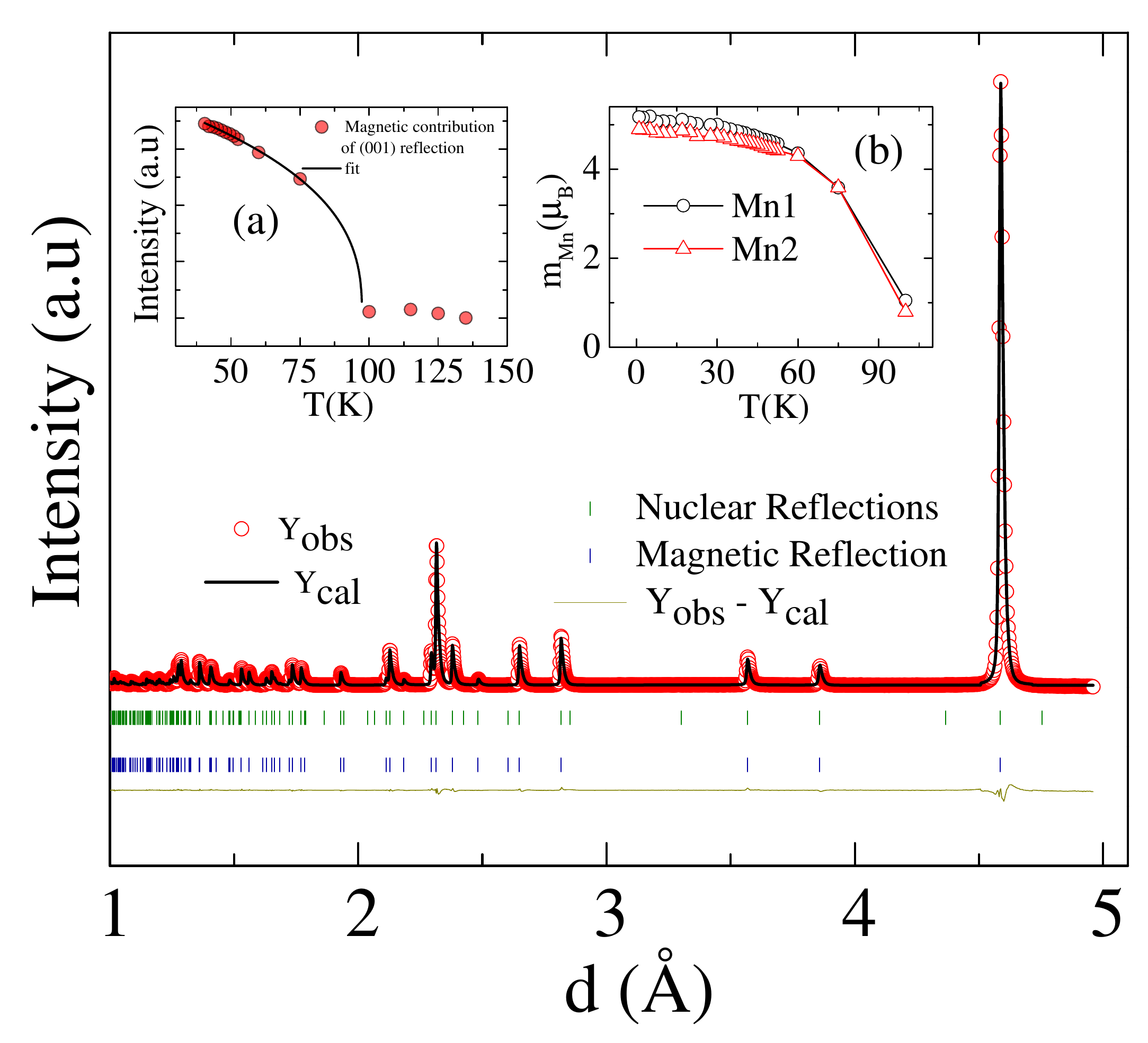}
  	\caption{ (Color online ) The  Rietveld refinement of the powder neutron diffraction data as obtained at 1 K, using Ireps $\Gamma_{2}$ . The inset (a) shows the temperature dependence of magnetic contribution to the intensity of the (001) reflection and solid line represents a fit. The inset (b) depicts the temperature dependence of the magnetic moments on both the crystallographic sites as deduced by refinement of neutron data.}
  	\label{Fig5}
  \end{figure}

Comparison of the neutron diffraction patterns recorded in the paramagnetic and the magnetically ordered states confirm that the magnetic Bragg contributions are restricted to the nuclear peak positions, with a pronounced enhancement of intensity being observed for the (100) and (200) reflections within the magnetically ordered state.  Temperature dependence of the magnetic intensity of (100) Bragg peak has been fit (inset of Fig.\ref{Fig5} (a)) by the equation $I=I_0 \left(1-\frac{T}{T_N}\right) ^{2\beta}$, and we obtain $T_N$ = 97.41 K and $\beta$ = 0.182.  This would suggest that the nature of magnetic interactions in Mn$_{4}$Ta$_{2}$O$_{9}$ is intermediate to that described by the two-dimensional (2D) ($\beta$ = 0.125) and three-dimensional (3D) ($\beta$ = 0.32) Ising models. We note that this is analogous to that reported for the magnetoelectrics LiCoPO$_4$ and Co$_4$Nb$_2$O$_9$ \cite{LiCoPO4,ArimaCoNbO}.
  
\begin{figure}
	\centering
	\includegraphics[scale=0.30]{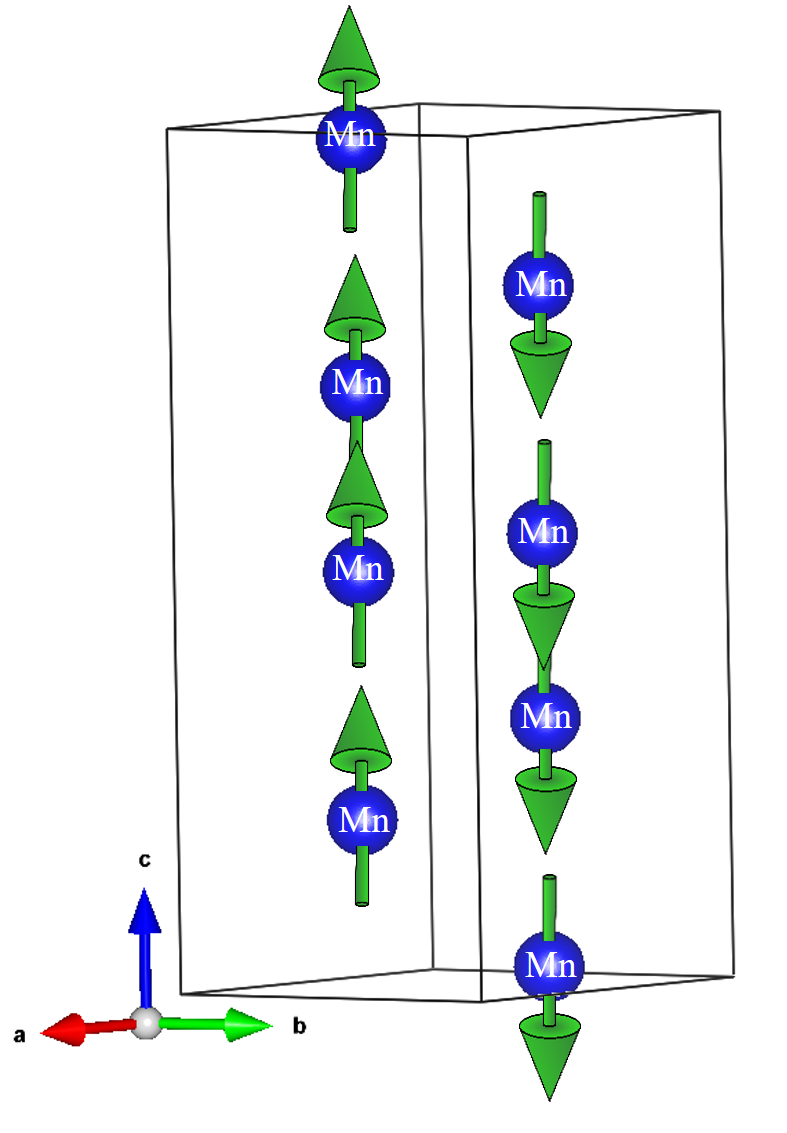}
	\caption{ (Color online ) (a) Magnetic structure of Mn$_4$Ta$_2$O$_9$ at 1K as obtained from the Rietveld refinement of T.O.F diffraction data with Ireps $\Gamma_{2}$. The structure comprises of antiferromagnetically coupled ferromagnetic chains of Mn${^{2+}}$ ions aligned along the trigonal $c$ axis.}
	\label{Fig6}
\end{figure}

\begin{figure}
	\centering
	\includegraphics[scale=0.25]{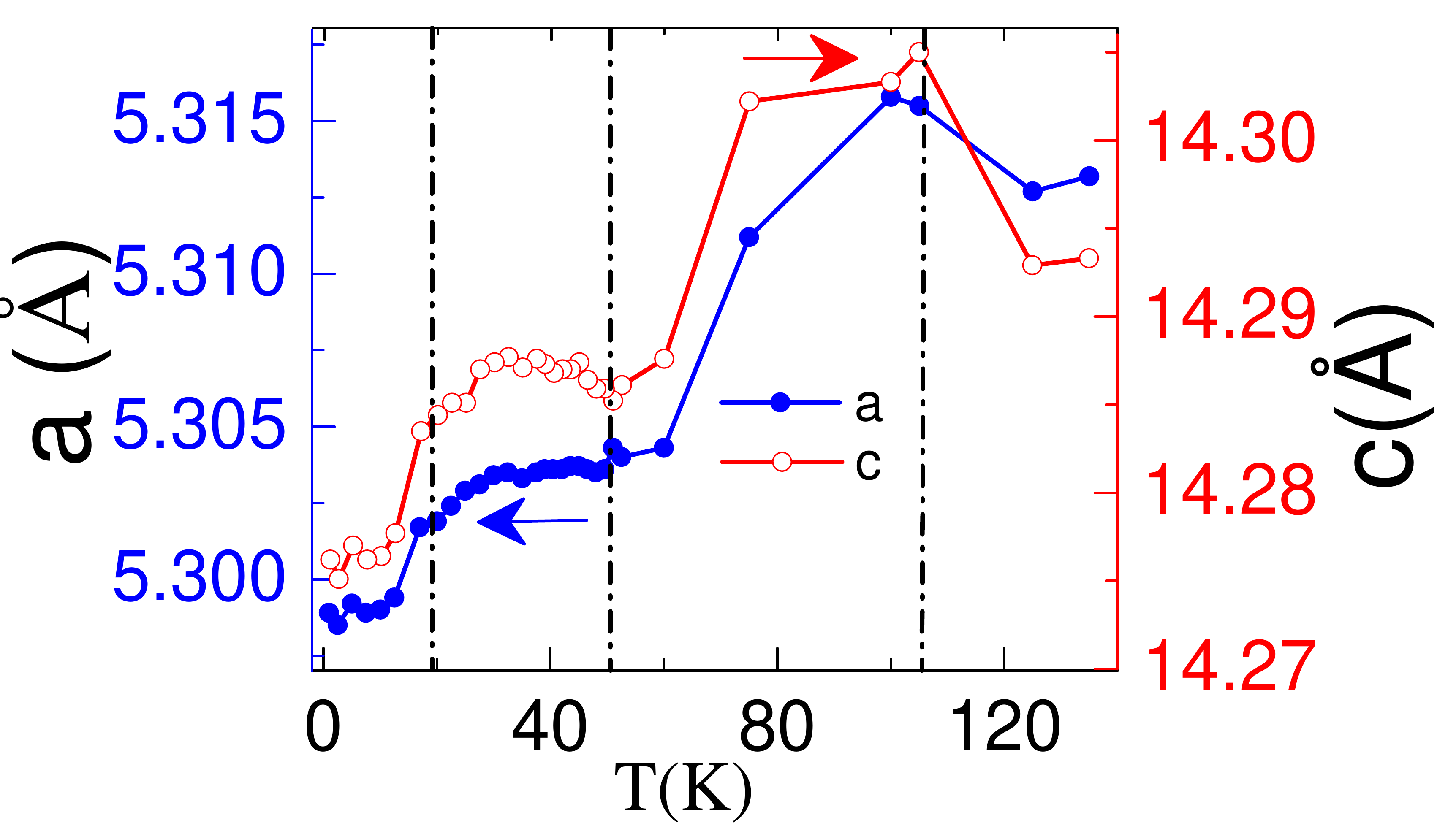}
	\caption{ (Color online ) Temperature variation of the lattice parameters $a$ (left axis) \& $c$ (right axis). The dash-dotted line indicates the temperatures of successive magnetic transitions as inferred from the magnetization measurements.}
	\label{Fig7}
\end{figure}

\begin{figure}
 	\centering
 	\includegraphics[scale=0.45]{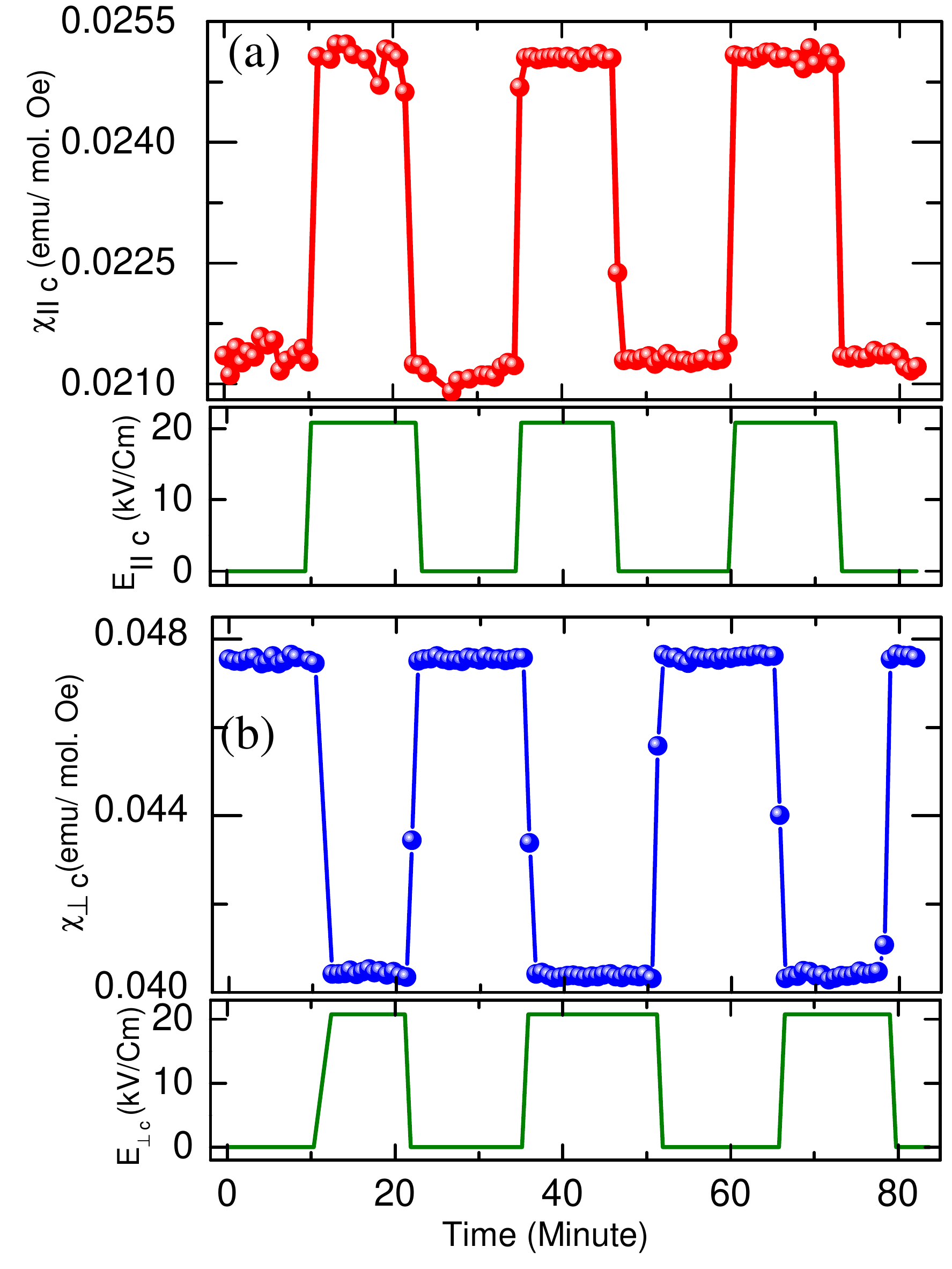}
 	\caption{ (Color online) Time dependent magnetic measurements of  Mn$_{4}$Ta$_{2}$O$_{9}$ under a periodic square electric pulse with (a) $H\parallel c$ and (b) $H\perp c$ performed at 40K. }
 	\label{Fig8}
 \end{figure}

The K search program in FULLPROF suggested that the most possible propagation vector (\bm {$\vec{K}$})  down to the lowest measured temperature is (000). For symmetry analysis, a combination of axial vectors localized on 4d (for Mn1 amd Mn2) have been considered and the basis functions of the irreducible representations of the wave vector group k = (0 0 0) were tested for symmetry. 
The decomposition of the magnetic structure contained six possible irreducible representations (Ireps)
\newline\hspace*{\fill}
$\Gamma_{mag}$=\bm{$\Gamma_{1}$}$\oplus$\bm{$\Gamma_{2}$}$\oplus$\bm{$\Gamma_{3}$}$\oplus$\bm{$\Gamma_{4}$}$\oplus$2\bm{$\Gamma_{5}$}$\oplus$2\bm{$\Gamma_{6}$}
\hspace*{\fill}\newline
Out of the six possible Ireps, $\Gamma_{1}$ - $\Gamma_{4}$ represents Mn$^{2+}$ moments aligned along the crystallographic $c$ axis, and $\Gamma_{5}$, $\Gamma_{6}$ represents the Mn$^{2+}$ aligned along the $ab$ plane \cite{SI}. Refinement with  $\Gamma_2 $ gave the best fit to the magnetic contribution of our diffraction data. The magnetic $R{_{Bragg}}$ factor was of the order of 1.7, with the next best fit (for $\Gamma_1 $) being more than ten times larger.  Rietveld refinement of the diffraction data at  1K with Ireps \bm{$\Gamma_2 $} is depicted in the main panel of Fig.\ref{Fig5}, and represents a ferromagnetic chain of  Mn spins coupled antiferromagnetically along the $c$ axis as shown in Fig.\ref{Fig6}.  We note that the magnetic structure which we have inferred is different to that described earlier for the iso-structural member Co$_4$Nb$_2$O$_9$ where the spins are aligned within the $ab$ plane with a canting either along the $c$ axis \cite{ArimaCoNbO} or in the $ab$ plane \cite{G_Deng}.  Assuming the spin only moment value with the orbital moment (L) fully quenched, the  calculated value of effective moment of (Mn$^{2+}$, d$^5$) in the low spin state ( S= 1/2 ) \textit{t$_{2g}^5$} \textit{$e_g^0$} is $\sqrt{3}$ = 1.73 $\mu_B$ and in high spin state ( S= 5/2 ) \textit{t$_{2g}^3$} \textit{$e_g^2$} is $\sqrt{35}$ = 5.91 $\mu_B$. The magnetic moments of Mn$^{2+}$ for Mn1 and Mn2 as obtained from the Reitveld refinement are 5.1(6) $\mu_{B}$ and 4.8(8) $\mu_{B}$ respectively as depicted in Fig. \ref{Fig5}(b), and they are close to the high-spin S = 5/2 state for Mn$^{2+}$. The inferred magnetic structure as well as the magnitude of the magnetic moment are in good agreement with a recent report\cite{xyz}.  The weak ferromagnetic component observed in our bulk magnetic measurement perpendicular to the crystallographic $c$ axis  is not clearly discernible from the powder neutron diffraction data, and hence our model does not account for the modulation in the magnetic structure which gives rise to this component. However, both the lattice parameters $a$ and $c$ exhibits discernible  anomalies across the observed magnetic transitions, as is depicted in Fig. \ref{Fig7}. \\           
The existence of magnetoelectric coupling is typically demonstrated by the means of magnetic field dependent dielectric and pyroelectric measurements, and it has now been demonstrated that Mn${_4}$Ta${_2}$O${_9}$ exhibits a finite electric polarization on the application of a magnetic field \cite{Fang}. The cross coupling between magnetic and polar orders can also be demonstrated by electric field ($E$) dependent magnetization ($M$) measurement, though this technique is relatively underutilized. We have performed time dependent magnetic measurements under a square wave electric field (of $20.8$kV/cm) in the $E\parallel c$ and the $E\perp c$ crystallographic orientations to monitor the change in magnetization as a function of applied electric fields. Here, the sample was first electromagnetically poled ($H\parallel E$, with $H$ = 2T and $E$ = 20.8kV/cm) down to 40K, after which the electric and magnetic fields were switched off. The magnetization was then measured at a constant magnetic field of 1000 Oe in the presence of a periodic square wave electric field of 20.8 kV/cm. The effect which this periodic electric field has on the magnetization along both the crystallographic axes is depicted in Fig.\ref{Fig8}.(a) \& (b). The magnetization is seen to be modulated as a function of the applied electric field, clearly reinforcing the presence of strong magnetoelectric coupling in Mn${_4}$Ta${_2}$O${_9}$. Moreover, this magnetoelectric coupling is also observed to be highly anisotropic, with the magnetization increasing on the application of the electric field along the $c$ axis, whereas it decreases when the electric field is applied perpendicular to it. The absence of experimental artifacts were confirmed by performing the same measurement on a slab of the insulating  paramagnet Er${_2}$O${_3}$, where no change in magnetization was observed on the application of an electric field. 
It is well established that the magnetoelectric effect in the prototypical magnetoelectric Cr$_2$O$_3$ itself is highly anisotropic \cite{Folen}. The coupling along the trigonal $c$ axis ($\alpha{_{\parallel}}$) and that perpendicular to it ($\alpha{_{\perp}}$) is reported to exhibit different temperature dependencies, with  $\alpha{_{\parallel}}$ also exhibiting a change of sign at $\approx$ 100K. In general, the magnetoelectric response in such antiferromagnetic lattices is ascribed to a combination of effects arising from the single ion anisotropy, the Lande $g$ factor, and the symmetric and antisymmetric magnetic exchange interactions \cite{Gehring}. The application of an electric field could have disparate effects on one or more of these contributory factors which are then reflected in the form of a change in magnetization. For instance, in the case of Cr$_2$O$_3$, it has been suggested that the temperature dependence in $\alpha{_{\parallel}}$ arises as a result of the modification of the intrasublattice magnetic exchange as a consequence of an applied electric field at higher temperatures, with the low temperature behavior being driven primarily by the electric field induced shift in $g$. On the other hand, the temperature dependence of $\alpha{_{\perp}}$ was ascribed to be dominated by the electric field dependence of the single ion anisotropy term \cite{Rado,Shtrikman}. 

The magnetoelectric response of the Co based members of this family have been investigated earlier, and it is reported that they exhibit a number of unique features \cite{HKusunose, G_Deng, ArimaCoNbO, solovyev, PRotatingH}. For instance it has been demonstrated that in Co${_4}$Nb${_2}$O${_9}$, the electric polarization rotates by an angle $-2\phi$ around the trigonal axis, when the magnetic field is swept by an angle $\phi$, and was attributed to a magnetic field driven rotation of the spins in the honeycomb lattice \cite{PRotatingH}. It has also been suggested that in Co${_4}$Nb${_2}$O${_9}$ and Co${_4}$Ta${_2}$O${_9}$, the observed electric polarization is a vector addition of the effective magnetoelectric contributions from two nonequivalent magnetic sublattices. These two contributions partially compensate each other, and hence the anisotropic and relatively smaller value of the magnetic field induced electric polarization \cite{solovyev}. The case of Mn${_2}$Ta${_2}$O${_9}$ is slightly different, since our data indicates that its magnetic structure (comprised of antiferromagnetically coupled ferromagnetic chains along the $c$ axis) is different from that of Cr${_2}$O${_3}$ as well as the Co based members of this family of honeycomb magnets. An added complication is the presence of a weak ferromagnetic phase below 50K which has not been observed in the other members of this family till date. The fact that we observe an electric field induced decrease in magnetization in the $H\perp c$ direction where this weak ferromagnetic component is observed, leads us to speculate that the our observations could be a possible consequence of the anisotropic response of this antisymmetric Dzyaloshinskii Moriya component on the application of an electric field. However, more comprehensive experiments, monitoring the temperature evolution of $\alpha{_{\parallel}}$ and $\alpha{_{\perp}}$, as well as the effect of electric fields on the crystallographic and magnetic structures would be required to understand the microscopic mechanisms responsible for the magnetoelectric response in Mn${_4}$Ta${_2}$O${_9}$. 

In summary, we report on the temperature dependent magnetization as measured on a single crystal of  the magneto electric Mn${_4}$Ta${_2}$O${_9}$, which suggest that the magnetic moments are aligned along the crystallographic $c$ axis. This is also confirmed by neutron diffraction measurements which indicates that the magnetic structure comprises of ferromagnetic chains of Mn$^{2+}$ spins aligned along the [001] axis, which is in contrast to that observed in the closely related isostructural Co${_4}$Nb${_2}$O${_9}$ system.  Time dependent magnetic measurements performed under a square electric pulse performed at 40K indicate that the magnetelectric response parallel and perpendicular to the trigonal $c$ axis have different signs. Our observations suggest that the Mn$_{4}$Ta${_2}$O${_9}$ system would make for an valuable addition in the quest for understanding the interplay between magnetism and electric polarisation in magnetoelectric honeycomb lattices. 
\section{ACKNOWLEDGMENT}
The authors acknowledge Surjeet Singh for extending experimental facilities and useful discussions. S.N.P. acknowledges IISER Pune for a fellowship. S.N. acknowledges DST India for support through grant no. SB/S2/CMP-048/2013 and funding support by the Department of Science and Technology (DST, Govt. of India) under the DST Nanomission Thematic Unit Program $(SR/NM/TP-13/2016)$. .
\bibliography{Bibliography}
\end{document}